\documentstyle[prl,aps,multicol]{revtex}
\input{epsf}
\begin{document}
\draft
\title{Relaxation process in a regime of quantum chaos}

\author{Giulio Casati$^{(a,b,c)}$, Giulio Maspero$^{(a,b)}$ and 
Dima L. Shepelyansky$^{(d,*)}$}
\address{$^{(a)}$Universit\`a di Milano, sede di Como, Via Lucini 3,
22100 Como, Italy}
\address{$^{(b)}$Istituto Nazionale di Fisica della Materia, 
Unit\`a di Milano, Via Celoria 16, 20133 Milano, Italy}
\address{$^{(c)}$Istituto Nazionale di Fisica Nucleare, Sezione di Milano,
Via Celoria 16, 20133 Milano, Italy} 
\address {$^{(d)}$ Laboratoire de Physique Quantique, UMR C5626 du CNRS, 
Universit\'e Paul Sabatier, F-31062 Toulouse, France}

\date{received 11 June, revised 11 October, 1997}
\maketitle
\begin{abstract}
We show that the quantum relaxation process in a classically chaotic
open dynamical system is characterized by a quantum relaxation time 
scale $t_q$.
This scale is much shorter than the Heisenberg time and much larger than
the Ehrenfest time: $t_q \propto g^\alpha$ where $g$ is the conductance
of the system and the exponent $\alpha$ is close to $1/2$. As a result,
quantum and classical decay probabilities remain close up to values
$P \sim \exp{ \left( -\sqrt{g} \right) }$ similarly to the case of 
open disordered systems.
\end{abstract}
\pacs{PACS numbers: 05.45.+b, 03.65.Sq}

\begin{multicols}{2}
\narrowtext


Recently there has been a considerable interest in 
the statistical properties of the poles of the S matrix 
in mesoscopic quantum dots coupled to conducting
metallic leads \cite{Fyod,Brouwer}.
The statistical properties of these poles determine the effective 
life time of
particles inside the dot and therefore are directly 
related to the conductance  
fluctuations and current relaxation inside the dot. In fact, the problem of 
current relaxation in diffusive mesoscopic samples connected to 
leads has been 
addressed long time ago \cite{Alt8788}. Recently the interest to this
problem was renewed and new effective methods based on 
the supersymmetry approach 
have been developed to study the problem in more detail \cite{Muzy}.
For quasi one-dimensional metallic samples the results of \cite{Alt8788,Muzy} 
predict that the current in the sample, being proportional to the probability
$P(t)$ to stay inside the sample, will decay, up to a very long time,
in an  exponential way according to the 
classical solution of diffusive equation which describes the 
electron dynamics
in disordered metallic samples: $P(t) \sim \exp{(-t/t_c)}.$
Here $t_c \sim t_D=N^2/D$ is the diffusion time for a system of size $N$ 
with diffusion coefficient $D$. 

According to \cite{Alt8788,Muzy} the strong
deviation of quantum probability $P_q$ from its classical value $P$
takes place only for $t > t_H$ where the quantum probability decays as 
$P_q(t)\sim \exp{(-g \ln^2({t/t_H}))}$. Here, $t_H=1/\Delta (\hbar=1)$
is the Heisenberg time, $\Delta$ is the level spacing inside the sample
and $g=t_H/t_c=E_c/\Delta$ is the conductance of the sample with
Thouless energy $E_c=1/t_c$. At time $t_H$, $\ln P_q(t_H)/\ln P(t_H) \sim 2$.
As it was pointed out recently \cite{mirlin}, less strong deviations
$(\ln (P_q(t_q)/P(t_q)) \sim 1)$ should take place at a shorter time 
$t_q \sim t_c \sqrt{g}$ due to weak localization corrections according 
to equations obtained in \cite{Muzy}.
Up to now these theoretical predictions for open systems
have not been  checked neither by numerical computations nor by laboratory
experiments. Also the above results are based on an ensemble averaging
over disorder and their validity for a quantum {\it dynamical} system which 
has {\it one fixed} classical limit is not evident.
The investigation of this problem is also interesting from
the viewpoint of semiclassical correspondence in a regime 
with exponentially fast spreading of narrow
wavepackets due to which the Ehrenfest 
time scale \cite{leshouches} is very short: $t_E \sim \ln{ N}/\Lambda$
where $\Lambda$ is the Liapunov exponent.

In this paper we study the quantum relaxation process in a 
dynamical model of quantum
chaos where diffusion is caused by the underlying classical chaotic
dynamics. This model, introduced in \cite{borgonovi},
 describes a kicked rotator with absorbing
boundary conditions (when the momentum is larger than some critical value).
This open system can be considered as 
a model of light trapped in a small liquid
droplet with a deformed boundary in which the rays, 
with orbital momentum less 
than some critical value, escape from the droplet because the refraction 
angle exceeds the critical value \cite{Stone}.

Contrary to the standard  kicked rotator  model 
\cite{leshouches} in which the matrix of the evolution operator 
is unitary, the absorption breaks the unitarity of the evolution matrix
so that all eigenvalues move inside the unit circle.
In other words, each eigenvalue can be written in the form $\lambda=
e^{-i\epsilon}=\exp{(-iE-\Gamma/2)}$ where $\Gamma$ 
characterizes the decay rate of the eigenstate. In this way absorption 
corresponds to ideal leads without reflections back to the sample.
A similar approach, in which coupling to continuum
was studied on the basis of non Hermitian Hamiltonians, has been developed
and widely used by Weidenm\"uller {\it et al.} 
(see for example \cite{Weiden}).

In our model the quantum evolution of the wavefunction is described
by the following quantum map:
\begin{eqnarray} 
\label{qmap}
\bar{\psi} = \hat{U} \psi = \hat{{\cal{P}}} e^{-iT\hat{n}^2/4} e^{-ik
\cos{\hat{\theta}}}
 e^{-iT\hat{n}^2/4} \psi,
\end{eqnarray}
where $\hat{{\cal{P}}}$ is a projection operator 
over quantum states $n$ in the
interval $(- N/2, N/2)$.
Here, the commutator is $[\hat{n},\hat{\theta}]=-i$ and the
classical limit corresponds to $k \rightarrow \infty$, 
$T \rightarrow 0$ while
the classical chaos parameter $K=kT$ remains constant.
In the classical limit the dynamics is described by the Chirikov standard map
\cite{leshouches}:
\begin{eqnarray} \label{cmap}
\bar{n} = n + k \sin{ \left[ \theta + { T n \over 2} \right]},
\bar{\theta} = \theta + {T \over 2} (n+\bar{n}).
\end{eqnarray}

For the classical computations, in analogy with the quantum model, 
all classical trajectories escaped from the
interval $(-N/2,N/2)$ are absorbed and never return back. 
Due to this absorption, in the regime of strong chaos ($K \gg 1$) 
with one chaotic component (no islands of stability), 
the classical probability to stay inside the interval $(-N/2,N/2)$
decays exponentially with time: $P(t) \sim \exp (-\gamma_c t)$.
The time scale $t_c = 1/\gamma_c \sim t_D$ is 
determined by the diffusion time $t_D$ required to reach the absorbing 
boundary from the center. 
Since the diffusion rate is $D = <\Delta n^2> / \Delta t
\sim k^2 /2$ then $\gamma_c=E_c \sim 1/t_D = {k^2 / N^2}$. 
In order to study the quantum 
relaxation we fixed the classical chaos parameter $K=7$ and the ratio
$N/k=4$. In this way the diffusion time $t_D$ is constant when 
$N \rightarrow \infty$ and this allows to investigate the semiclassical 
behavior. Moreover $t_D >> 1$ which justifies the diffusive approximation.
We note also that the system (\ref{qmap}) with $-N/2 < n < N/2$ and
$\hat{{\cal{P}}} = \hat{{ \openone }}$ coupled to open leads ($T=0$ for
$\left| n \right| > N/2$) had been studied in \cite{italo}.
The results obtained there showed that this model has universal conductance 
fluctuations \cite{UCF} and other properties very similar to mesoscopic 
metallic samples.

The numerical solution of the classical problem was obtained by iterating
map (\ref{cmap}) for $M=9 \cdot 10^9$ different initial conditions
homogeneously distributed on the line $n=0$.
The results
demonstrate a clear exponential decay $P(t)=\exp(-\gamma_c t -b)$ 
with $\gamma_c=0.101882(1),  b=0.17774(5)$ (see Fig.1).
This exponential decay
shows that for $K=7$ the phase space is completely chaotic without 
any island of stability. Even with such a high number of orbits,
the classical computations allow to 
obtain {\it directly} the probability $P(t)$ 
with 10 \% accuracy only up to the level 
$\overline{P} \approx 5 \cdot 10^{-8} (\overline{t} \approx 165)$. 
This limitation is
due to statistical errors appearing for finite number of trajectories. 
In spite of this, the decay rate $\gamma_c$ can be found with very 
high precision which allows to extrapolate
the probability behaviour to larger times.

For the quantum evolution we choose the corresponding initial condition 
in which only the level $n=0$ is populated and 
we studied numerically the quantum dynamics  (\ref{qmap}) for different $N$.
We have found that
the quantum probability $P_q(t)$ follows the classical one 
up to a  time $t_q$
after which it starts to decay at a slower rate (Fig.1).
We determined the quantum relaxation time $t_q$ by 
the condition $\ln{ \left[P_q(t_q)/P(t_q) \right]} =0.1$ 
which corresponds to 10\% deviation.
The comparison of quantum and classical probabilities is shown in Fig.1.
The values of $t_q$, obtained in this way, strongly fluctuate with 
changing the system size $N$ as is typical for mesoscopic systems (Fig.2).
These fluctuations are satisfactory described by a log normal distribution
(insert in Fig.2), but a more detailed analysis is required to determine
precisely their statistical properties.

To suppress the fluctuations, we average $\ln{ P_q(t)}$ 
over different system
sizes by changing $N$ in a small interval $\delta N$ around a given $N$.
Typically we averaged up to $500$ different $N$ values. This allows
to determine the averaged ratio $R(t)= \left< \ln{ (P_q(t)/P(t))} \right>$ 
of quantum to classical probability.
For all these N values
the classical dynamics is {\it exactly the same} 
since we kept $K=7$ and $N/k=4$.
Then the quantum relaxation time $t_q$ (averaged) at 10\% level, is 
determined by the condition $R(t_q)=0.1$. 
The dependence of $R(t)$ on time, for different $N$, is shown in Fig.3.
It is clearly seen that $t_q$ grows as we approach the
semiclassical limit $N \rightarrow \infty$.
The rescaling of data for $R(t)$ as a function of $t/t_q$ shows
a satisfactory global scaling behaviour of quantum probability
(see insert in Fig.3).
The dependence of $t_q$ on $N$ is shown in Fig. 4 for the semiclassical
regime $500 < N \leq 130001$. This regime corresponds to a variation
of conductance $g=N/t_c$ in the interval $50 < g \leq 13000$.
The fit of numerical data gives a power law dependence
$t_q \approx 0.19 \cdot t_c \cdot N^\alpha$ with $\alpha \approx 0.41$.
This power remains the same for 5\% deviation level (Fig.4).

Here we propose a qualitative explanation of $\alpha$ value based on the fact
that in the open system
the physics is affected  not by 
the level spacing $\Delta$ but by the distribution of poles of 
the scattering matrix S which describes the coupling to the leads.
These poles are located in the complex energy plane and their imaginary parts
determine the decay probability of eigenmodes inside the sample.
For our model, the poles are simply given by the evolution operator
(\ref{qmap}).
The eigenvalues of $\hat{U}$ are distributed in a narrow ring of width
$E_c$ inside the unitary circle \cite{borgonovi}. This is typical
for diffusive samples coupled to strongly absorbing leads.
As a result, $N$ complex eigenvalues are homogeneously distributed in a ring
of total area $A \approx E_c$ and the distance between them, in 
the complex plane, is $\delta \approx \sqrt{ {E_c / N}}$.
In the classical limit this spacing goes to zero and one obtains a continuous
density of poles. However, for finite $N$, the separation 
of poles is finite and can be resolved after a time $t_q \sim 1 / \delta$. 
According to this argument, which is independent of the symmetry and 
dimensionality of the problem, the 
deviation between quantum and classical probabilities will take place at
\begin{eqnarray} \label{scal}
 t_q \approx 0.38 \sqrt{ t_c N} = 0.38 t_c \sqrt{g} 
\end{eqnarray}
where the numerical coefficient has been extracted from Fig.4.
The theoretical dependence  (\ref{scal}),
which corresponds to $\alpha=1/2$, is different but close to the numerical
value $\alpha=0.41$. We attribute this difference to a not sufficiently large 
value of $\sqrt{g}$. Indeed, neglecting the values with $\sqrt{g} < 15$, we 
obtain $\alpha = 0.44$ for 10 \% deviation and $\alpha=0.45$ 
for 5\% deviation;
these values are closer to the theoretical prediction $\alpha=0.5$.

The scale $t_q$ can be also explained in a more standard way 
based on weak localization corrections \cite{mirlin,Muzy}.
Indeed, the quantum interference gives a decrease of the diffusion rate
$1/t_c \propto D \rightarrow D(1- a t/t_H)$ where $a$ is some
constant (diffusion stops at time $t_H$). As a result
$\ln(P_q/P) \approx a t_q^2/(t_c t_H) \sim 1$ and one gets (3).

For very large times, the decay of $P_q(t)$ is determined by the eigenvalue
$\epsilon = E-i \Gamma/2$ with minimal $\Gamma=\Gamma_{min}$.
This asymptotic behaviour should start from some time scale $t_f$,
which can be estimated in a following way.
Similarly 
to the results obtained for complex matrices \cite{Felix},
the eigenvalues $\epsilon$ should be distributed in a region with smooth
boundary in the complex plane with approximately 
constant density of points for $\Gamma \sim \gamma_c$.
Typically this boundary is parabolic near the extremal $\Gamma_0$ close to 
$\Gamma_{min}$. Due to this, the relative number of eigenvalues
$dW$ in the interval $d\Gamma$ is given by 
$dW/d\Gamma \sim \sqrt{ (\Gamma -\Gamma_0)} /\gamma_c^{3/2}$.
The total probability in the interval $\delta \Gamma = \Gamma - \Gamma_0$ is
$W \sim (\delta \Gamma / \gamma_c )^{3/2}$. The distance between the two 
lowest values of $\Gamma$ can be estimated from the condition $W \sim 1/N$
which gives $\delta \Gamma \sim \gamma_c N^{-2/3}$.
Then $t_f \approx 0.4 t_c N^{2/3}$ which is much larger than $t_q$. The
numerical factor here was determined from the two lowest values of 
$\Gamma$ for $N=5001$: $\Gamma_{min} = \Gamma_1 = 0.065309$ 
and nearest $\Gamma_2 = 0.066203$. 
These values were obtained by direct diagonalization of the matrix $U$.
The rate of the asymptotic decay 
of $P_q(t)$ for $N=5001$ (Fig.1)
agrees, up to $6$ digits accuracy, with the above value of $\Gamma_{min}$.
We note that a typical size of fluctuations for poles is
$\delta$ and so we expect that $\gamma_c-\Gamma_{min} \sim \delta \sim 1/t_q$.
Since in our model $t_f \ll t_H$ and 
the classical limit is fixed we do not see
the quantum deviations discussed in \cite{Alt8788,Muzy} for $t > t_H$.
At large $g$ one should average over exponentially large number of realizations
to observe them.

The largest value of $t_q$ we have numerically obtained (at $N=1.3 \cdot 10^5$)
is $t_q=254$ which corresponds to a probability $P_q(t_q) \approx
5 \cdot 10^{-12}$. On the other hand, the classical simulation with 
$M=9\cdot 10^9$ orbits allows to directly compute the classical
$P(t)$ with 10\% accuracy only up to $t \approx 165$ which corresponds
to a probability $\overline{P} \approx 5\cdot 10^{-8}$.
In order to reach the above level of quantum accuracy one should iterate
$M \approx 10^{14}$ orbits, which is already out of the possibility of present
day computers. Moreover, the value of $P_q$ at 10 \% accuracy level can be
easily increased 
by orders of magnitude since according to (\ref{scal}) $t_q$ grows
proportionally to $\sqrt{N}$. Instead,  for classical computations,
the number of required orbits $M$ will increase exponentially 
($\overline{P} \sim 1/M$).
This demonstrates that quantum computations of exponential relaxation
processes are much more efficient than classical simulations with large 
number of orbits.
The possibility to efficiently compute the quantum probability $P_q(t)$
up to very long times $t_q >> 1/\Lambda$, allows to numerically estimate 
the measure of the 
integrable component $\mu$ in the phase space of the classical system.
Indeed, for $\mu > 0$, the existence of integrable islands leads to an
asymptotic power law decay of correlations
$P(t) \propto t^{-0.5}$ \cite{chir2}.
Since in our numerical data the quantum probability decays exponentially
up to $P_q(t_q)\approx 5 \cdot 10^{-12}$ we assume that the measure 
of the integrable component is $\mu < P_q(t_q) 
\approx 5 \cdot 10^{-12} \times (1 \pm 0.1)$
being much smaller than the relative size of quantum cell $1/N$.
Here, the error bar gives the average fluctuation of $P_q(t_q=254)$ 
obtained for 77 values of $N$.

Also it is interesting to note that 
the Ehrenfest time scale $t_E$ is much smaller than the quantum relaxation
time $t_q$:
$t_E/t_q \sim \ln{N} / \sqrt{N} << 1$. For example 
for $N=1.3 \cdot 10^5$ we have $t_q=254$ while
$t_E = \ln{N} / \Lambda 
\approx 9.4 $ ($\Lambda \approx \ln{ K/2} \approx 1.25$).
This shows that the agreement between quantum and classical relaxation
continues for a time scale which is much larger than the time of
wave packet spreading.
However, for $t > t_E$ there is no exponential instability
in the quantum motion \cite{leshouches,dls}. As a result, 
correlation functions of the type 
$C(\tau)=\left< \sin \theta(t) \sin \theta(t+\tau) \right>$
which, in the regime of strong chaos, decay exponentially 
in the classical case 
 ($\ln \left| C \right| \sim - \Lambda \tau)$, in the quantum case decay
only during the Ehrenfest time scale up to 
$\ln \left| C \right| \sim - \ln N \;\; (t_E \ll \tau \ll t_c)$. 
This is similar to what happens in closed (unitary) systems such 
as the kicked rotator \cite{dls}. This example shows that exponential 
relaxation is not necessary related to exponential local instability
and positive Kolmogorov-Sinai entropy.

After the submission of this paper the scale $t_q$ has been
obtained on the basis of random matrix theory and
supersymmetry for kicked rotator with random phases \cite{SavFra}.
The related results for pre-localized states in closed systems 
were discussed in \cite{Falko}.

\pagebreak 

\begin{figure}
\centerline{\epsfxsize=10cm \epsfbox{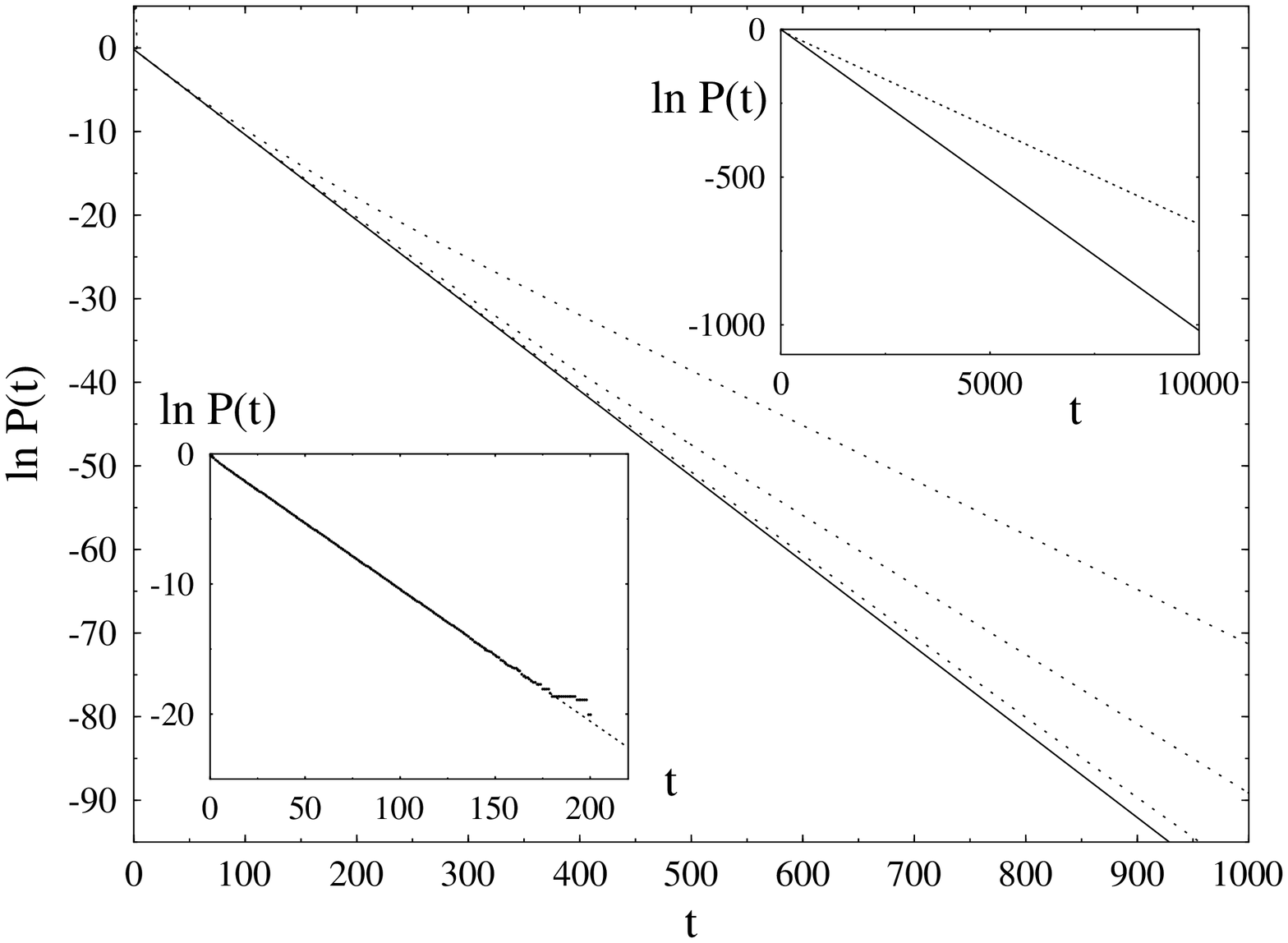}}
\vspace{4mm}                                   
\caption{Classical and quantum probability decay for $K=7$ and $N/k=4$.
The full line shows the fit to the classical decay. Dotted lines give 
the quantum probability $P_q(t)$ for $N=5001, 20001, 130001$ (upper, middle
and lower curves respectively). 
The lower insert shows the classical probability actually computed 
from $M=9 \cdot 10^9$ orbits and the fit is shown by the 
dotted line. The upper insert shows the classical (full line) 
and the quantum (dotted line) asymptotic decay for $N=5001$.  }
\end{figure}

\begin{figure}
\centerline{\epsfxsize=10cm \epsfbox{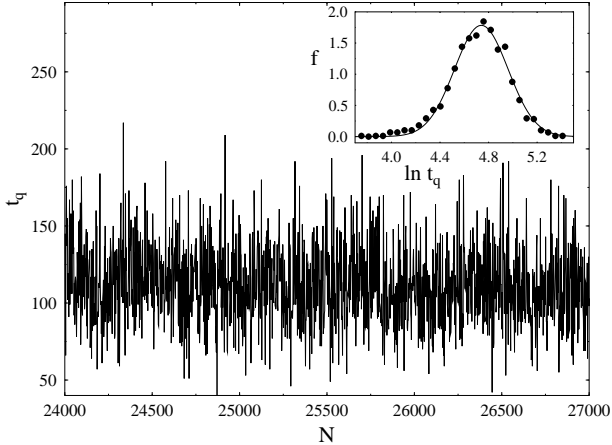}}
\vspace{4mm}                                   
\caption{Mesoscopic fluctuations of the quantum relaxation time
$t_q$ for different system sizes $N$. The insert shows the statistical
distribution of fluctuations f which is close to a log normal 
distribution of width $0.22$.}
\end{figure}

\begin{figure}
\centerline{\epsfxsize=10cm \epsfbox{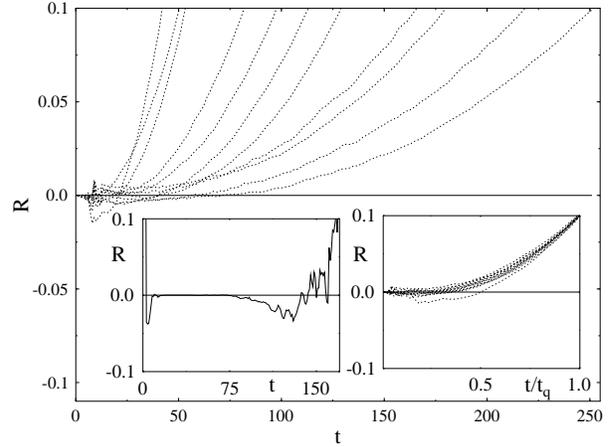}}
\vspace{4mm}                                   
\caption{The average ratio $R(t) = \left< \ln{ \left( P_q(t) / P(t) \right) }
\right>$ as a function of time for different $N$ from $N=2500$ (left curve)
to $N=129500$ (right curve). The horizontal full line
corresponds to $P_q(t)=P(t)$. The probability $P(t)$ is given by
numerical data obtained with $M=9 \cdot 10^9$ orbits for $t \leq 70$ and
by the fit from Fig.1 (see text) for $t > 70$.
The left insert shows the ratio of the 
numerically computed classical probability  $P(t)$
to the fit function. The deviations from the fit, for $t > 70$, are
due to statistical errors related to finite $M$.
The right insert demonstrates the scaling behaviour of $P_q(t)$ on the variable
$t/t_q$ where the $t_q$ values are determined by condition $R(t_q)=0.1$.}
\end{figure}

\begin{figure}
\centerline{\epsfxsize=10cm \epsfbox{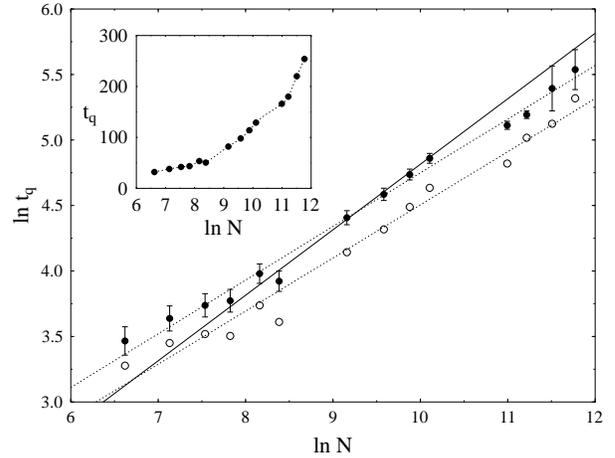}}
\vspace{4mm}                                   
\caption{Dependence of the quantum relaxation time $t_q$ on the system
size $N=g t_c$ in logarithmic scale. Points refers to 10 \% deviation
level ($R=0.1$) while circles refer to 5\% deviation ($R=0.05$).
The two dotted lines give the fit $t_q = 1.9 N^{0.41}$ and
$t_q=1.5 N^{0.41}$ respectively.
The full line gives the theoretical prediction with $\alpha=1/2$
(\ref{scal}).
The insert shows the 10 \% data in a semilog scale.}
\end{figure}

\end{multicols}

\end{document}